\documentclass[aps,pra,showpacs,twocolumn,floatfix]{revtex4-1}
\usepackage[utf8]{inputenc}
\usepackage{graphicx,times}
\usepackage{amsmath}
\usepackage{amsfonts}
\usepackage{braket}

\usepackage{hyperref}

\usepackage[final,authormarkup=none]{changes}
\definechangesauthor[name={Matteo Rossi}, color=red]{MR}

\newcommand{\expected}[1]{\left\langle#1\right\rangle}
\newcommand{\id}{\mathbb{I}}
\newcommand{\hamiltonian}{\mathcal{H}}
\newcommand{\mx}{\text{\tiny max}}
\newcommand{\R}{\text{R}}
\newcommand{\nn}{\text{\tiny N}}
\newcommand{\sep}{\text{\tiny{SEP}}}
\newcommand{\ghz}{\text{\tiny{GHZ}}}
\newcommand{\rnd}{\text{\tiny{RND}}}
 
\newcommand{\ie}{\text{\tiny{IE}}}
\newcommand{\ce}{\text{\tiny{CE}}}

\begin{document}
\title{Entangled quantum probes for dynamical environmental noise}
\author{Matteo A. C. Rossi}\email{matteo.rossi@unimi.it}
\affiliation{Dipartimento di Fisica, Universit\`a
degli Studi di Milano, 20133 Milano, Italy}
\author{Matteo G. A. Paris}\email{matteo.paris@fisica.unimi.it}
\affiliation{Dipartimento di Fisica, Universit\`a 
degli Studi di Milano, 20133 Milano, Italy}
\date{July 15, 2015}
\begin{abstract}
We address the use of entangled qubits as quantum probes to characterize
the noise induced by complex environments. In particular, we
show that \added[id=MR]{a joint measurement on }entangled probes can improve estimation of the correlation time for a broad class of environmental noises compared to any sequential strategy involving single qubit preparation. The enhancement appears
when the noise is faster than a threshold value, a regime which may
always be achieved by tuning the coupling between the quantum probe and the
environment inducing the noise. Our scheme exploits time-dependent
sensitivity of quantum systems to decoherence and does not require
dynamical control on the probes. We derive the optimal interaction time 
and the optimal probe preparation, showing that it corresponds 
to multiqubit GHZ states when entanglement is useful. 
We also show robustness of the scheme against depolarization or dephasing
of the probe, and discuss simple measurements approaching optimal 
precision.
\end{abstract}
\pacs{03.67.-a, 05.40.-a, 03.65.Yz}
\maketitle
The coherence properties of a quantum system are strongly
affected by its interaction with the surrounding environment. This is
often an obstacle to the implementation of quantum technologies, so 
that much effort has been devoted to study the system-environment 
interaction and to engineer decoherence in order to minimize its 
degrading effects \cite{Davies1976,Breuer2002}. 
On the other hand, the very sensitivity of quantum 
systems to external influence also provides an effective tool to 
characterize unknown parameters of a given environment 
\cite{Escher2011,Alipour2014} by exploiting {\em quantum
probes}, as opposed to classical ones, usually macroscopic and more
intrusive.
Indeed, characterizing the noise induced by an external
complex system is of great relevance in many areas of nanotechnology, as
well as in monitoring biological or chemical processes
\cite{Hanggi1995,Taylor2008,Mittermaier2009,Zhong2001}. Besides, it
represents a crucial step to design robust quantum protocols resilient
to noise \cite{Bylander2011,Zhang2007,AliAhmed2013,
Almog2011,Banchi2014,Vasile2014}. 
\par
The proper framework to address characterization by 
quantum probes \cite{Chin2012,Haikka2014}, and 
to design the best working conditions, is given by quantum estimation 
theory \cite{Paris2009}, which provides analytical tools to optimize the three 
building blocks of an estimation strategy: (i) preparation of the probe 
system in a suitably optimized state, (ii) controlled interaction of the 
probe with the system for an optimal amount of time $t$, (iii) 
measurement of an optimal observable on the probe. Overall, the 
ultimate precision for any unbiased estimator $\hat \gamma$ of 
a certain parameter $\gamma$ is bounded by the quantum 
Cram\`er-Rao (CR) theorem, stating that $\text{Var}(\hat \gamma) 
\geq [M H(\gamma)]^{-1}$, where $M$ is the number of measurements 
and $H(\gamma)$ is the quantum Fisher information (QFI), i.e. 
the superior of the Fisher information over all possible quantum 
measurements described by positive operator-valued measures (POVMs).
\par
Recently, single-qubit quantum probes have been proposed for the 
characterization of noise by monitoring decoherence and 
dephasing induced by the environment under 
investigation, in particular when the latter can be described in terms of classical stochastic processes \cite{Alvarez2011,Paris2014,Benedetti2014a,Cywinski2014,
Benedetti2014b}. Indeed, 
stochastic modeling of the environment has been proven \cite{Crow2014} to reliably describe noise and 
the decoherence process in several systems affected by dephasing
\cite{Neder2011,Biercuk2011,Witzel2014,Fink2014,Yu2010,Li2011,Benedetti2013a,Benedetti2012}. It may also be useful for several other systems of interest, 
including motional averaging \cite{Li2013} and solid state qubits 
\cite{Burkard2009,Wold2012,Benedetti2014c}.
\par
In this paper, we extend this analysis to entangled qubits used 
as quantum probes, and show how they greatly improve 
the characterization of a broad class of 
environmental noises compared to any sequential strategy 
involving single qubit preparation \cite{DAriano2001,Fujiwara2001,Acin2001}. 
In particular, we show how to improve estimation of the correlation time 
(i.e. the spectral width) of classical noise. Since such noise is usually emerging from a large 
collection of fluctuators, we are going to consider Gaussian stochastic processes.
\begin{figure}[h!]
\includegraphics[width=0.9\columnwidth]{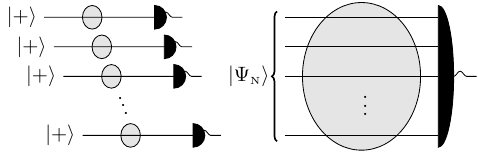}
\caption{Schematic diagram of possible characterization techniques. 
On the left: we 
have $N$ qubits and each one is prepared in the state $\ket{+}$, then 
interacts with
the system for a time $t$ and is finally measured independently of 
the other qubits. On the right: the $N$ qubits are initially prepared 
in a multiqubit GHZ state $\ket{\Psi_\nn}$, and are let
interact with the system such that they are subject to the same
realization of the noise. At the output, a collective measurement is
performed on the qubits.} \label{fig:scheme}
\end{figure}
\par 
The probing scheme is depicted in Fig.~\ref{fig:scheme}, both
for a sequence of $N$ uncorrelated qubits and for an $N$-qubit entangled
state. In both cases we assume that the qubits do not interact 
with each other. In each case, the qubits may interact with different realizations of the noise or with the same realization, depending on the temporal and spatial distance between the probes. We end up with four possible schemes, which we describe in detail in the Appendix. In the following, we focus on the best case for each configuration, i.e. independent realizations for the separable probes and a common environment for entangled probes.
\par
Let us start by considering a single qubit interacting 
with a fluctuating dephasing environment. The Hamiltonian is given by
$
\hamiltonian(t) = \omega_0 \sigma_z + B(t) \sigma_z,
$
where $\omega_0$ is the energy of the qubit and $B(t)$ is a realization 
of the stochastic process that describes the noise. As a 
paradigmatic example we consider a zero-mean 
Ornstein-Uhlenbeck process characterized by the autocorrelation 
function $K(t,t') \equiv \braket{B(t)B(t')}_B = \frac12 
\gamma\Gamma\exp[-\gamma |t-t'|]$, or by the corresponding 
Lorentzian spectrum. 
Here, $\gamma$ is the spectral width, i.e. the inverse of the
autocorrelation time, while $\Gamma$ denotes the coupling between the
probe and the system.  It is worth noticing that a similar analysis may
be carried out for other Gaussian processes, e.g. processes with
power-law or Gaussian 
autocorrelation functions, and that results are qualitatively 
the same, independently on the choice of the autocorrelation function.
\par
The density operator of the evolved qubit is
given by 
\begin{equation} \rho(t) =
\braket{U(t)\rho(0)U^\dagger(t)}_B,
\end{equation}
where $\braket{\cdot}_B$ denotes the average over all
possible realizations of the stochastic process in the time interval
$[0,t]$, $U(t) = \exp[-i \int_0^t \hamiltonian(s) ds]= \exp\{[-i
\omega_0 t + \phi(t)]\sigma_z\}$ is the time evolution operator, and
$\phi(t) = \int_0^t B(s)ds$ is the accumulated phase due to the
environmental (dynamical) noise.
An explicit expression for $\rho(t)$ can be found by employing
the characteristic function of a zero-mean Gaussian stochastic 
process:
$
\expected{e^{i m \phi(t)}}=e^{-\frac 12 m^2 \beta_\gamma(t)},
$
where 
\begin{equation}\label{eq:beta}
	\beta_\gamma(t) = \int_0^t\int_0^t ds\, dw\, K(s,w) = 
\frac{\Gamma}{\gamma} (e^{-\gamma  t}+\gamma  t-1).
\end{equation}
\par
If the qubit is initially prepared in a state described by 
the density operator $\rho(0)$, the density operator at the 
time $t$ will be $\rho(t)$ with $\rho_{kk}(t) = 
\rho_{kk}(0)$, $k=1,2$ and 
\begin{equation}
\rho_{12} (t)
= e^{-2\left[i\omega_0 t+\beta_\gamma(t)\right]}\,\rho_{12}(0)\,.
\end{equation}
The optimal single qubit preparation, given by 
$|+\rangle =\frac1{\sqrt{2}} (|0\rangle + |1\rangle\added[id=MR]{)}$, 
together with the corresponding
QFI and the optimal measurement for the estimation of
$\gamma$ have recently been found 
\cite{Benedetti2014b}. For $N$
uncorrelated qubits, thanks to additivity, the QFI is just $N$ times 
the single qubit QFI, i.e. 
\begin{equation}\label{eq:sep_QFI} 
H_\nn^{\sep}(\gamma,t) = \frac {4 N}{e^{4
\beta_\gamma(t)}-1} [\partial_\gamma \beta_\gamma(t)]^2.  
\end{equation}

Let us now consider a probe made of $N$ qubits initially prepared 
in the generalized GHZ entangled state $
	\ket{\Psi_\nn} = (\ket{0}^{\otimes N} 
	+ \ket{1}^{\otimes N})/\sqrt{2}\,$,
\replaced[id=MR]{interacting with a common environment}{We also assume that the qubits do not interact with each other 
and that they interact with a common environment, i.e. with 
the same realization of the noise}.The overall Hamiltonian is thus
\begin{equation}\label{eq:entangled_hamiltonian}
   \hamiltonian^{(\nn)}(t) = \hamiltonian(t) \otimes 
   \id ^{\otimes N-1} + \id \otimes 
   \hamiltonian(t) \otimes \id^{\otimes N-2} + \ldots,
\end{equation}
where $\hamiltonian(t)$ is the above single qubit Hamiltonian
and $B(t)$ is the same realization of the noise for all the 
qubits. 
The QFI for the parameter $\gamma$ reads
\begin{equation}\label{eq:ghz_QFI}
H_\nn^{\ghz}(\gamma,t) = 
\frac{4 N^4}{e^{4 N^2 \beta (t,\gamma )}-1} 
[\partial_\gamma \beta_\gamma(t)]^2\,\replaced[id=MR]{.}{,}
\end{equation}

\begin{figure}
	\includegraphics[width=.99\columnwidth]{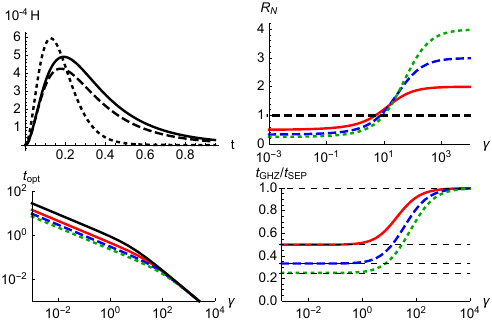}
\caption{(Color online) In the upper left panel, the QFI 
$H_2^\sep(t)$ (solid) and $H_2^\ghz(t)$ (dotted) as function of time for $\gamma = 10$. We also show for comparison the (smaller) QFI for separable two-qubit probes in a common environment (dashed) . In the upper right the QFI ratio $\R_\nn$ as a function of $\gamma$ (log scale) for $N=2$ (solid red), $N=3$ (dashed blue) and $N=4$ (dotted green). For small values of $\gamma$, the ratio 
is below one (black dashed line), and tends asymptotically 
to $1/N$: in this regime it is more convenient to employ 
separable states than maximally entangled states.
The ratio increases monotonically with $\gamma$ and exceeds one at a
threshold value $\gamma_0(N)$, which depends on $N$. For asymptotically
large $\gamma$, the use of $N$-qubit GHZ states is $N$ times better than
the use of $N$ qubits in a separable state.
In the lower left pannel, a log-log plot of the optimal interaction time as a function of $\gamma$ for separable probes (black), and for an entangled probe with $2$ (solid red), $3$ (dashed blue) and $4$ (dotted green) qubits. In the lower right panel, the ratio between the optimal time for the GHZ state, $t_\ghz$ and for the separable state $t_\sep$.} 
\label{fig:ratio_ghz}
\end{figure}

\replaced[id=MR]{We notice that $\beta_\gamma(t)$ is a 
monotonically increasing function of $t$ with $\beta_\gamma(0) = 0$ (see Eq.~\eqref{eq:beta})}{where $\beta_\gamma(t)$ (see Eq.~\eqref{eq:beta}) is a 
monotonically increasing function of $t$ with $\beta_\gamma(0) = 0$}. 
Moreover, we have $\beta_\gamma(t) \sim
\Gamma(t-1/\gamma)$ for $t\gg 1$. Thus both $H_\nn^\sep (\gamma,t)$ and
$H_\nn^{\ghz}(\gamma,t)$ are asymptotically vanishing and show a single 
maximum, corresponding to different optimal values of the interaction 
time, $t_{\text{opt}}^\sep$ and $t_{\text{opt}}^\ghz$ respectively.
We refer to this maximum value as the maximal QFI for a specific value 
of $\gamma$. As is apparent from the above equations, the maximization 
of the QFI involves transcendental equations and must be done numerically. 
The behavior of $H_\nn^\sep(\gamma,t)$ and $H_\nn^{\ghz}(\gamma,t)$ is 
depicted in the upper left panel of Fig.~\ref{fig:ratio_ghz}. The lower panels of the figure show how the optimal time depends on $\gamma$ for the two measurement schemes.

\par
In situations where we can control the interaction time between the 
probe and the environment, it will be most convenient to set it to the 
optimal time. Thus, a fair comparison between separable and entangled
probes naturally involves the maximal QFI of the two cases. We therefore 
introduce the QFI ratio as
$\R_{ \nn}(\gamma) = {H^\ghz_{\nn,\mx}(\gamma)}/{H^\sep_{\nn,
\mx}(\gamma)}\,$,
and analyze its behavior as a function of $\gamma$ and $\text{N}$.
When $\R_\nn(\gamma)>1$, the use of a $N$ qubits in a GHZ state improves 
estimation compared to the use of $N$ uncorrelated probes, e.g. in a
sequential strategy.  Figure \ref{fig:ratio_ghz} illustrates the main
results: the ratio $\R_\nn(\gamma)$ is larger than one for $\gamma > 
\gamma_0(N)$, where $\gamma_0(N)$ is a threshold value that depends 
on $N$. Moreover, $\R_\nn(\gamma) \rightarrow N$ for $\gamma \gg 
\gamma_0(N)$. This result is enhanced by the fact that, upon 
substituting $\tilde \gamma = \gamma / \Gamma$ and $\tau = \tilde \gamma t$, 
we may show that the quantum signal-to-noise ratio (QSNR) $\tilde 
\gamma^2 H(\tilde \gamma)$ does not depend on $\Gamma$. This means that, 
if one is able to control the coupling between the probe and the system, 
one can always tune $\Gamma$ to achieve a situation where 
$\R_\nn(\tilde \gamma) > 1$. All the Figures are obtained by setting
$\Gamma=1$. 
\par
Now a question arises: Is the GHZ state $\ket{\Psi_\nn}$ the optimal
one? Are there (entangled) states that give even higher QFI? The answer 
to this question cannot be analytic, because one cannot diagonalize 
analytically a generic density matrix of a multiqubit state. In order
to attack this problem, we first notice that the maximum is achieved 
for an initial pure state \cite{Fujiwara2001}. 
We have thus generated a large number ($n= 10^6$) of random initial 
pure states, uniformly distributed according to the Haar measure, 
for different values of $\gamma$ and for $N = 2,3,4$. For each random 
state, the maximal QFI, $H^\rnd_\nn (\gamma)$, resulting from the interaction
with a common environment has been numerically evaluated using 
the expression
$H(\gamma) = 2 \sum_{nm} {|\braket{\psi_m|\partial_\gamma\rho_\gamma|\psi_n}|^2}/(\rho_n+\rho_m),
$
where $\rho_\gamma = \sum_n \rho_n \ket{\psi_n}\bra{\psi_n}$ is 
the diagonal form of the density operator after the interaction with
the environment. This value is then used to evaluate the corresponding 
QFI ratio $H^\rnd_\nn (\gamma)/H^\sep_\nn (\gamma)$, and 
to compare the estimation precision to the precision achievable 
using $N$ independent qubits 
interacting with separate environments.
Our results
indicate that, for $\gamma \gtrsim \gamma_0$, that is, in the 
region where entanglement is convenient, the GHZ state is indeed the 
optimal one, thus showing that entanglement is a resource for the 
estimation of the spectral width of Gaussian noise.
\par
Below the threshold the GHZ state interacting with a common environment is no longer optimal and the optimal strategy involves separable probes interacting with independent environments. For completeness, we anyway look for the optimal state in a common environment and found numerically  that the extremal state lies in the same family of states that had been 
identified in \cite{Huelga1997} as optimal probes
to improve frequency estimation.
The states of this family, for $N$
qubits, have the form 
$
\ket{\Phi_\nn} = \sum_{k=0}^{\lfloor \frac12 \nn \rfloor} a_k \ket{k},
$
where $a_k$ are normalized real coefficients, $\ket{k}$ is an equally
weighted superposition of all $N$-qubit states with a number $k$ or a
number $N-k$ of excitations, and $\lfloor \cdot \rfloor$ denotes the
integer part.  The GHZ state belongs
to this family with $a_0 = 1/\sqrt{2}$ and all other coefficients
set to $0$. 
\begin{figure}[h!]  
\includegraphics[width=0.9\columnwidth]{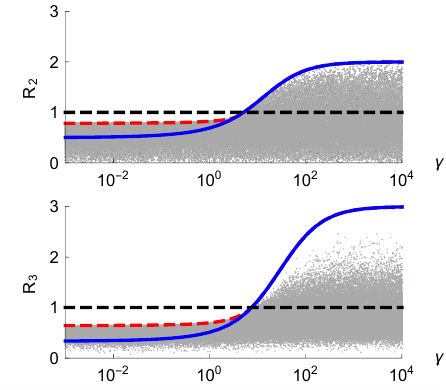}
\caption{(Color online) The curves in the logplots show the  
QFI ratio $\R_2$ (top) and $\R_3$ (bottom) as a function of 
$\gamma$ for the GHZ state (solid blue) and the optimal state of 
$N$ qubits in a common environment  (dashed red). The two curves 
superimpose each other above the threshold $\gamma_0 (N)$. We also 
show the QFI ratio for $10^5$ randomly generated states (gray points), 
uniformly distributed according to the Haar measure.}
\label{fig:optimal_random_ghz_ratio} 
\end{figure}
\par
Figure~\ref{fig:optimal_random_ghz_ratio} illustrates our numerical 
results obtained for two and three qubits. The plots show the QFI 
ratios $\R_2$ (left) and $\R_3$ (right). The solid blue line is the 
ratio for the GHZ state, the $10^4$ gray points correspond to the QFI
ratio of randomly generated states and the dashed red line is found by
optimizing the QFI over the coefficients $a_k$ of $\ket{\Phi_\nn}$. 
We can see that from $\gamma \gtrsim
\gamma_0(N)$ the blue and red curves coincide, i.e. GHZ states are
extremal.
We also notice that for $\gamma \gtrsim \gamma_0(N)$ a significant
fraction of gray dots lies above the $r=1$ dashed line, but the dots are
sparse around the solid blue line, meaning that the GHZ state allows for a
remarkable gain in the estimation of larger values of $\gamma$.
\par
\begin{figure}[h!]
\includegraphics[width=0.8\columnwidth]{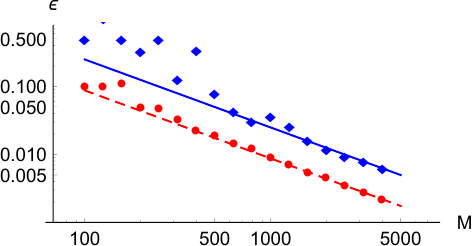} 
\caption{(Color online) Log-log plot of the 
relative error $\epsilon = \sqrt{\text{Var}(\hat\gamma)}/\hat\gamma$
of the Bayesian estimator 
as a function of the number of
measurements, for $\gamma = 10$. The lines represent the CR bound
for a single-qubit measurement (dashed blue) and for a 4-qubit GHZ 
state (solid red). The diamonds (single qubit) and the dots (4 qubits) 
correspond to the performance of a Bayes estimator applied to 
simulated experiments. Bayes estimators saturate the
CR bound when increasing the number of measurements and very good 
performances are achieved already for thousands of measurements.
The plot also shows that estimation improves with the number of 
qubits since the CR bound is saturated with a lower number of 
measurement.} \label{fig:bayes}
\end{figure}

It is worth to emphasize that optimal precision, i.e. the QFI of 
Eq. \eqref{eq:ghz_QFI} may be achieved upon implementing a simple 
rank-2 measurement. Indeed, $H_\nn^\ghz(\gamma)$ corresponds to 
the Fisher information of a projective measurement on the two 
eigenvectors corresponding to the nonzero eigenvalues of the evolved  
density operator, which are, respectively
\begin{align}
\label{eq:ghz_eigenvalues}
p_{\pm} &= \frac{1}{2} \left(1\pm e^{-2 N^2 
\lambda ^2 \beta_{\gamma }(t)}\right) \\
\label{eq:ghz_eigenvectors}
\ket{p_\pm} & = \frac 1 {\sqrt 2}  
(e^{- i \omega_0 \sigma_z})^{\otimes N}
\left(\ket{0}^{\otimes N} \pm \ket{1}^{\otimes N}\right).
\end{align}
The CR theorem, however, sets only a {\em lower bound} to the precision 
of any unbiased estimator, and a question arises on how to suitably 
process data coming from the above rank-2 measurement in order to {\em
saturate} the bound. Bayesian estimators are known to saturate the CR 
bound for asymptotically large numbers of measurements: in order to  
assess quantitatively the performance of Bayesian 
estimation we have  performed simulated experiments on the probing 
system. In particular, we have simulated the outcomes 
$\{x_1,\ldots,x_M\}$ of the measurement by
randomly choosing a result according to the probabilities of Eq.
\eqref{eq:ghz_eigenvalues} and have built a Bayesian estimator $\hat
\gamma$ as the mean value of the {\em a posteriori distribution}, 
starting from a flat prior. 
The resulting relative error of $\hat \gamma$, $\epsilon 
= \sqrt{\text{Var}(\hat\gamma)}/\hat\gamma$, is shown, as a function of
the number of measurements, in Fig.  \ref{fig:bayes} for a specific
value of $\gamma$. We see that with a relatively low number of measurements,
of the order of thousands, the bound is saturated and the situation
improves by increasing the number of qubits. The proposed scheme
thus allows for an effective and achievable estimation of the parameter
$\gamma$.
\par
\begin{figure}[h!]
\includegraphics[width=0.9\columnwidth]{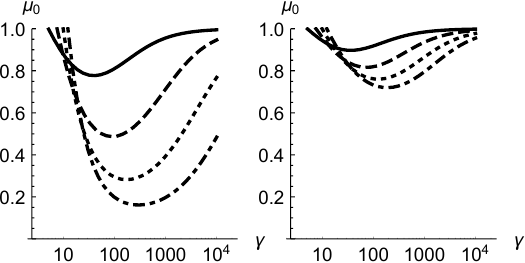}
\caption{Threshold purity $\mu_0$, as a function of the spectral width 
$\gamma$, above which the use of a 
depolarized (left panel) or a dephased GHZ state 
(right panel) is more effective than a set of uncorrelated 
qubits. The different
lines correspond to a different number $N$ of qubits: $N=2$ (solid),
$N=3$ (dashed), $N=4$ (dotted), $N=5$ (dot-dashed). The threshold 
$\mu_0$ approaches 1 when $\gamma \rightarrow \gamma_0(N)$ and when
$\gamma \rightarrow \infty$, whereas there is an intermediate region
where $\mu_0$ decreases to a minimum, meaning that there is more
tolerance in the initial preparation of the probe. When $N$
increases the minimum of $\mu_0$ decreases and moves 
to larger values of $\gamma$.} \label{fig:threshold_purity}
\end{figure}

Let us now address the robustness of our scheme against noise in the preparation of the probe. In fact, we have shown that the use of entangled qubit probes 
prepared in a GHZ state leads to enhanced precision in the 
estimation of the spectral width. However, it is
generally challenging to experimentally 
prepare the probes \emph{exactly} in the 
GHZ state and a question arises on how sensible is this estimation 
scheme to, e.g., the purity of the initial preparation. We answer this question 
by considering a partially depolarized state, 
$\rho_p =  p\, \rho_{\ghz} + (1-p)\id/2^N$, where $\id$ is the 
identity matrix and $0<p<1$, and a partially dephased state, 
$\rho_\delta = \delta \rho_\ghz+ \frac12 
(1-\delta)(\ket{0}^{\otimes N}\bra{0}+\ket{1}^{\otimes N}\bra{1})$ 
where $0< \delta < 1$. In both cases, an analytic expression for 
the QFI may be found: we have
\begin{align}\label{eq:depolarized_QFI}
H^p_\nn(\gamma,t) &= \frac{2^{N+2} N^4 \left[\left(2^{N}-2\right) 
p+2\right]p^2 [\partial_\gamma \beta_\gamma(t)]^2}
{\left[\left(2^N-2\right) p+2\right]^2 e^{4 N^2 
\beta (t,\gamma )}-4^N p^2}\\
H_\nn^{\delta}(\gamma,t) &= 
\frac{4 N^4 \delta^2}{e^{4 N^2 \beta (t,\gamma )}-\delta^2} 
[\partial_\gamma \beta_\gamma(t)]^2\,,
\end{align}
which are obviously less than $H^\ghz_\nn(\gamma)$, being $\rho_p$ 
and $\rho_\delta$ mixed states, but may be still larger than 
$H_\nn^\sep(\gamma)$. Indeed, Figure~\ref{fig:threshold_purity} 
shows that for each value of $\gamma$ above $\gamma_0(N)$ there 
is a threshold value for the purity, above which the use of a 
depolarized or dephased GHZ state still leads to an improvement 
over the use of $N$ uncorrelated probes. The threshold 
purity $\mu_0$ is close to one for $\gamma \simeq \gamma_0(N)$ and 
for $\gamma \gg \gamma_0(N)$, whereas it shows a minimum in the 
intermediate region, thus allowing for a certain tolerance in the 
preparation of 
the initial state of the probe. Besides, this minimum value of the
threshold gets lower when increasing the number of qubits. 
\par
In conclusion, we have shown that the use of entangled qubits 
as quantum probes outperforms the sequential use of single-qubit 
probes in the characterization of the noise induced 
by complex environments. In particular, we have shown that \added[id=MR]{a joint measurement on} entangled 
probes improves estimation of the correlation time
for a broad class of environmental noises 
when the noise is faster than a threshold value.
This result is enhanced by
the fact that, upon controlling the coupling between the probe and the
system, the threshold value can be reduced arbitrarily.
Our scheme exploits time-dependent
sensitivity of quantum systems to decoherence and does not require
dynamical control on the probes. We have found the optimal 
interaction time 
and the optimal multiqubit probe preparation, showing that it corresponds 
to multiqubit GHZ states.
The proposed measurement scheme achieves the
Cram\'er-Rao bound for a relatively low number of 
measurements, upon employing a Bayesian estimator, and
is robust against imperfect preparation of the initial entangled state.
\begin{acknowledgments}
	This work has been supported by MIUR through the FIRB project
	``LiCHIS'' (grant RBFR10YQ3H), by EU through the Collaborative 
	Project QuProCS (Grant 
	Agreement 641277) and by UniMI through the H2020 Transition 
	Grant 15-6-3008000-625. MGAP thanks Claudia Benedetti and Rafal 
	Demkowicz-Dobrzanski for discussions.
\end{acknowledgments}

\appendix
\section{Supplemental Material}
We consider probes prepared both in the separable state $\ket{+}^{\otimes N}$ and in the generalized GHZ entangled state $
\ket{\Psi_\nn} = (\ket{0}^{\otimes N} 
+ \ket{1}^{\otimes N})/\sqrt{2}\,$.

We also consider two possible scenarios: in the first, each qubits interacts with an independent realization of the noise. This means that the overall Hamiltonian is 
\begin{equation}
   \hamiltonian^{(\nn)}(t) = \hamiltonian_1(t) \otimes 
   \id ^{\otimes N-1} + \id \otimes 
   \hamiltonian_2(t) \otimes \id^{\otimes N-2} + \ldots,
\end{equation}
where the realizations of the stochastic processes in each Hamiltonian $\hamiltonian_i(t)$ are uncorrelated, and the expected value of Eq. (2) must be calculated over all possible realizations of $B_1(t),\ldots,B_N(t)$.
In the second scenario all the qubits interact with a common environment and the qubits interact with the same realization of the noise. Then $\hamiltonian_1(t) = \ldots =\hamiltonian_N(t) = \hamiltonian(t)$ and the expected value in Eq. (2) must be calculated over all possible realizations of a single stochastic process $B(t)$.

We now show the results involving all the four possible probing schemes and show that a probing scheme involving $N$ qubits initially prepared in a GHZ state and interacting with a common environment outperforms any probing scheme involving $N$ qubits in the separable state $\ket{+}^{\otimes N}$ when $\gamma$ is greater than a threshold value.

\subsection{Separable probes, independent enviroments}
Since each qubit interacts with an independent realization of the noise, this scheme amounts to $N$ repetitions of the measurement of a single qubit probe prepared in the optimal state, Eq. (24) of Ref. \cite{Benedetti2014b}, and thus, thanks to the additivity of the QFI,
\begin{equation}\label{eq:sep_ie_QFI} 
H_\nn^{\sep,\ie}(\gamma,t) = \frac {4 N}{e^{4
\beta_\gamma(t)}-1} [\partial_\gamma \beta_\gamma(t)]^2.  
\end{equation}

\subsection{Separable probes, common environment}
\label{par:product_probe_common_environment}
In this scenario the dynamics of each qubits is not independent and we need to determine the dynamics of the whole $N$-qubit state. The QFI has a readable analytical form only for two qubits
\begin{equation}
H_2^{\sep,\ce}(\gamma,t)=\frac{32 \left\{e^{8 \beta_\gamma(t)} [\sinh 4 \beta_\gamma(t)+1]+1\right\}}{3 e^{16 \beta_\gamma(t)}-2 e^{8 \beta_\gamma(t)}+1}[\partial_\gamma \beta_\gamma(t)]^2.
\end{equation}

One can easily see that
\begin{equation}
\forall t,\gamma\qquad H_2^{\sep,\ie}(\gamma,t) > H_2^{\sep,\ce}(\gamma,t),
\end{equation}
since $\beta_\gamma(t) > 0$. 

\begin{figure}
\includegraphics[width=.8\columnwidth]{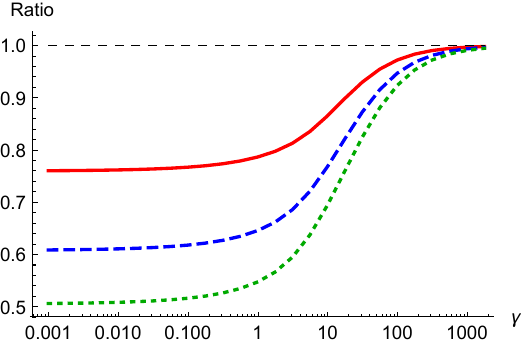}
\caption{Ratio between $H_\nn^{\sep,\ce}(\gamma,t)$ and $H_\nn^{\sep,\ie}(\gamma,t)$ for $N = 2$ (solid red), $N = 3$ (dashed blue), $N = 4$ (dotted green). The ratio is always lower than one and gets lower as $N$ increases. In the limit $\gamma \gg 1$ the ratio reaches one.}
\label{fig:QFI_Ratio_sep}
\end{figure}

We  checked numerically that the maximal QFI for separable probes interacting with a common environment is always lower to the maximal QFI for separable probes interacting with independent environments for fixed $\gamma$ also for $N = 3$ and $N = 4$. The results, shown in Fig. \ref{fig:QFI_Ratio_sep}, indicate that the ratio between $H_n^{\sep,\ce}(\gamma,t)$ and $H_2^{\sep,\ie}(\gamma,t)$ decreases with $N$. 

\begin{figure}
\includegraphics[width=.7\columnwidth]{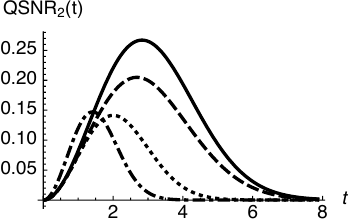}\\
\includegraphics[width=.7\columnwidth]{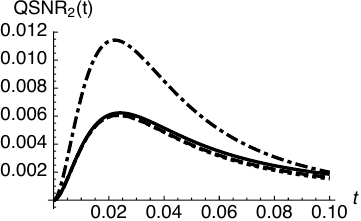}
\caption{The two plots show the two-qubit quantum signal-to-noise ratio (QSNR) $\gamma^2 H(\gamma,t)$ as a function of time for two values of the spectral width $\gamma$ of the noise: on the left $\gamma = 0.1$ and on the right $\gamma = 100$. The solid line shows $\gamma^2 H_\nn^{\sep,\ie}(\gamma,t)$, the dashed line $\gamma^2 H_\nn^{\sep,\ce}(\gamma,t)$, the dotted line $\gamma^2 H_\nn^{\ghz,\ie}(\gamma,t)$ and the dot-dashed line $\gamma^2 H_\nn^{\ghz,\ce}(\gamma,t)$. From the left panel we can see that $H_\nn^{\sep,\ie}(\gamma,t) > H_\nn^{\sep,\ce}(\gamma,t)$ for all $t$ and that $H_{\nn,\max}^{\ghz,\ce}(0.1) > H_{\nn,\max}^{\ghz,\ie}(0.1)$. From the right panel we can see that for high values of $\gamma$ the GHZ probes interacting with a common environment outperforms the other schemes.}
\label{fig:qfi_vs_t}
\end{figure}

\begin{figure}
\includegraphics[width=0.8\columnwidth]{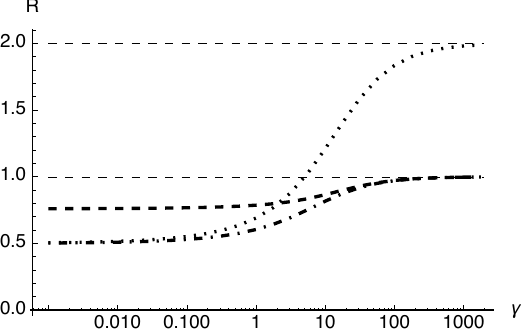}
\caption{Ratios between $H_{2,\max}^{\sep,\ce}(\gamma)$ (dashed),  $H_{2,\max}^{\ghz,\ie}(\gamma)$ (dotted) $H_{2,\max}^{\ghz,\ce}(\gamma)$ (dot-dashed) and $H_{2,\max}^{\sep,\ie}(\gamma)$ as functions of $\gamma$. We can see that the GHZ probes interacting with common environment achieves a higher maximal QFI than the other schemes. Similar plots can be produced for $N > 2$.}
\label{fig:ratios}
\end{figure}

\subsection{GHZ probes, independent environments}
In this case, the expected value of the density operator at time $t$ over al possible realizations of the stochastic processes $B_1(t), \ldots, B_N(t)$, is 
\begin{equation}
\begin{split}
\rho_\ghz(t) = & \frac 12 (\ket{0\ldots 0}\bra{0\ldots 0} + \ket{1\ldots 1}\bra{1\ldots 1}) + \\
& \frac{1}{2} e^{-2 N \beta _{\gamma }(t)}(e^{-2 i N \omega_0}\ket{0\ldots 0}\bra{1\ldots 1} +  \text{h.c.})
\end{split}
\end{equation}
and we find, for the QFI,
\begin{equation}\label{eq:ghz_ie_QFI} 
H_\nn^{\ghz,\ie}(\gamma,t) = \frac{4 N^2}{e^{4 N \beta_\gamma (t)}-1}[\partial_\gamma\beta_\gamma(t)]^2.  
\end{equation}
It is quite easy to prove that $H_\nn^{\ghz,\ie}(\gamma,t) < H_\nn^{\sep,\ie}(\gamma,t)$ for all $t$ and $\gamma$ and so we don't have an improvement in the estimation of $\gamma$ with a probe in the GHZ state if each qubit interacts with an independent realization of the environment.

\subsection{GHZ probes, common environment}
If the entangled qubits of the probe are affected by the same realization of the noise one finds that the expected value of $\rho(t)$ is
\begin{equation*}
\begin{split}
\rho_\ghz(t) = & \frac 12 (\ket{0\ldots 0}\bra{0\ldots 0} + \ket{1\ldots 1}\bra{1\ldots 1}) + \\
& \frac{1}{2} (e^{-2 N i \omega_0 - 2 N^2 \beta _{\gamma }(t)} \ket{0\ldots 0}\bra{1\ldots 1} +  \text{h.c.})
\end{split}
\end{equation*}
and obtains the following expression for the QFI:
\begin{equation}
H_\nn^{\ghz,\ce}(\gamma,t) = 
\frac{4 N^4}{e^{4 N^2 \beta (t,\gamma )}-1} 
[\partial_\gamma \beta_\gamma(t)]^2.
\end{equation}

\subsection{Maximal QFI values} 
\label{sub:maximal_qfi_values}
We have shown the analytical equations for the four measurement schemes. 
Figure \ref{fig:qfi_vs_t} shows the dependence on time of the QFI in the four cases and for two values of $\gamma$, which are respectively well below and well above the threshold value $\gamma_0$. 

We can see that the QFI as a function of time has one maximum. The position of the maximum may not be found analytically, due to the transcendental nature of the optimization equations, but can be found numerically with arbitrary precision.

In Fig. \ref{fig:ratios} we show the ratios between the maximal values of the QFI for the various cases and the maximal value for the QFI for separable probes interacting with an independent environment, as functions of $\gamma$, in the two-qubit case. We can see that the the ratios are below one except for the scheme involving a joint measurement on entalged probes that interact with a common environment, when $\gamma > \gamma_0$. Analogous plots may be produced for $N > 2$.

\bibliography{library.bib}

\begin{thebibliography}{39}%
\makeatletter
\providecommand \@ifxundefined [1]{%
 \@ifx{#1\undefined}
}%
\providecommand \@ifnum [1]{%
 \ifnum #1\expandafter \@firstoftwo
 \else \expandafter \@secondoftwo
 \fi
}%
\providecommand \@ifx [1]{%
 \ifx #1\expandafter \@firstoftwo
 \else \expandafter \@secondoftwo
 \fi
}%
\providecommand \natexlab [1]{#1}%
\providecommand \enquote  [1]{``#1''}%
\providecommand \bibnamefont  [1]{#1}%
\providecommand \bibfnamefont [1]{#1}%
\providecommand \citenamefont [1]{#1}%
\providecommand \href@noop [0]{\@secondoftwo}%
\providecommand \href [0]{\begingroup \@sanitize@url \@href}%
\providecommand \@href[1]{\@@startlink{#1}\@@href}%
\providecommand \@@href[1]{\endgroup#1\@@endlink}%
\providecommand \@sanitize@url [0]{\catcode `\\12\catcode `\$12\catcode
  `\&12\catcode `\#12\catcode `\^12\catcode `\_12\catcode `\%12\relax}%
\providecommand \@@startlink[1]{}%
\providecommand \@@endlink[0]{}%
\providecommand \url  [0]{\begingroup\@sanitize@url \@url }%
\providecommand \@url [1]{\endgroup\@href {#1}{\urlprefix }}%
\providecommand \urlprefix  [0]{URL }%
\providecommand \Eprint [0]{\href }%
\providecommand \doibase [0]{http://dx.doi.org/}%
\providecommand \selectlanguage [0]{\@gobble}%
\providecommand \bibinfo  [0]{\@secondoftwo}%
\providecommand \bibfield  [0]{\@secondoftwo}%
\providecommand \translation [1]{[#1]}%
\providecommand \BibitemOpen [0]{}%
\providecommand \bibitemStop [0]{}%
\providecommand \bibitemNoStop [0]{.\EOS\space}%
\providecommand \EOS [0]{\spacefactor3000\relax}%
\providecommand \BibitemShut  [1]{\csname bibitem#1\endcsname}%
\let\auto@bib@innerbib\@empty
\bibitem [{\citenamefont {Davies}(1976)}]{Davies1976}%
  \BibitemOpen
  \bibfield  {author} {\bibinfo {author} {\bibfnamefont {E.~B.}\ \bibnamefont
  {Davies}},\ }\href {https://books.google.com/books?id=nsh-AAAAIAAJ{\&}pgis=1}
  {\emph {\bibinfo {title} {{Quantum theory of open systems}}}}\ (\bibinfo
  {publisher} {Academic Press},\ \bibinfo {year} {1976})\BibitemShut {NoStop}%
\bibitem [{\citenamefont {Breuer}\ and\ \citenamefont
  {Petruccione}(2007)}]{Breuer2002}%
  \BibitemOpen
  \bibfield  {author} {\bibinfo {author} {\bibfnamefont {H.-P.}\ \bibnamefont
  {Breuer}}\ and\ \bibinfo {author} {\bibfnamefont {F.}~\bibnamefont
  {Petruccione}},\ }\href@noop {} {\emph {\bibinfo {title} {{The Theory of Open
  Quantum Systems}}}}\ (\bibinfo  {publisher} {Oxford University Press},\
  \bibinfo {year} {2007})\BibitemShut {NoStop}%
\bibitem [{\citenamefont {Escher}\ \emph {et~al.}(2011)\citenamefont {Escher},
  \citenamefont {{de Matos Filho}},\ and\ \citenamefont
  {Davidovich}}]{Escher2011}%
  \BibitemOpen
  \bibfield  {author} {\bibinfo {author} {\bibfnamefont {B.~M.}\ \bibnamefont
  {Escher}}, \bibinfo {author} {\bibfnamefont {R.~L.}\ \bibnamefont {{de Matos
  Filho}}}, \ and\ \bibinfo {author} {\bibfnamefont {L.}~\bibnamefont
  {Davidovich}},\ }\href {\doibase 10.1038/nphys1958} {\bibfield  {journal}
  {\bibinfo  {journal} {Nat. Phys.}\ }\textbf {\bibinfo {volume} {7}},\
  \bibinfo {pages} {406} (\bibinfo {year} {2011})}\BibitemShut {NoStop}%
\bibitem [{\citenamefont {Alipour}\ \emph {et~al.}(2014)\citenamefont
  {Alipour}, \citenamefont {Mehboudi},\ and\ \citenamefont
  {Rezakhani}}]{Alipour2014}%
  \BibitemOpen
  \bibfield  {author} {\bibinfo {author} {\bibfnamefont {S.}~\bibnamefont
  {Alipour}}, \bibinfo {author} {\bibfnamefont {M.}~\bibnamefont {Mehboudi}}, \
  and\ \bibinfo {author} {\bibfnamefont {A.}~\bibnamefont {Rezakhani}},\ }\href
  {\doibase 10.1103/PhysRevLett.112.120405} {\bibfield  {journal} {\bibinfo
  {journal} {Phys. Rev. Lett.}\ }\textbf {\bibinfo {volume} {112}},\ \bibinfo
  {pages} {120405} (\bibinfo {year} {2014})}\BibitemShut {NoStop}%
\bibitem [{\citenamefont {H{\"{a}}nggi}\ and\ \citenamefont
  {Jung}(1995)}]{Hanggi1995}%
  \BibitemOpen
  \bibfield  {author} {\bibinfo {author} {\bibfnamefont {P.}~\bibnamefont
  {H{\"{a}}nggi}}\ and\ \bibinfo {author} {\bibfnamefont {P.}~\bibnamefont
  {Jung}},\ }in\ \href
  {https://books.google.it/books?hl=it{\&}lr={\&}id=KyjaLJU7jQoC{\&}oi=fnd{\&}pg=PA239{\&}dq=P.+H{\%}C3{\%}A4nggi+and+P.+Jung+Colored+Noise+in+Dynamical+Systems++Advances+in+Chemical+Physics+89,+239{\%}E2{\%}80{\%}93326+(1995){\&}ots=iyXBKVcaLe{\&}sig={\_}cldi2nNI-2oK0QJd1PlvbiVHl8}
  {\emph {\bibinfo {booktitle} {Adv. Chem. Phys.}}},\ Vol.~\bibinfo {volume}
  {89},\ \bibinfo {editor} {edited by\ \bibinfo {editor} {\bibfnamefont
  {I.}~\bibnamefont {Prigogine}}\ and\ \bibinfo {editor} {\bibfnamefont
  {S.~A.}\ \bibnamefont {Rice}}}\ (\bibinfo  {publisher} {John Wiley {\&} Sons,
  Inc.},\ \bibinfo {year} {1995})\ pp.\ \bibinfo {pages} {239--326}\BibitemShut
  {NoStop}%
\bibitem [{\citenamefont {Taylor}\ \emph {et~al.}(2008)\citenamefont {Taylor},
  \citenamefont {Cappellaro}, \citenamefont {Childress}, \citenamefont {Jiang},
  \citenamefont {Budker}, \citenamefont {Hemmer}, \citenamefont {Yacoby},
  \citenamefont {Walsworth},\ and\ \citenamefont {Lukin}}]{Taylor2008}%
  \BibitemOpen
  \bibfield  {author} {\bibinfo {author} {\bibfnamefont {J.~M.}\ \bibnamefont
  {Taylor}}, \bibinfo {author} {\bibfnamefont {P.}~\bibnamefont {Cappellaro}},
  \bibinfo {author} {\bibfnamefont {L.}~\bibnamefont {Childress}}, \bibinfo
  {author} {\bibfnamefont {L.}~\bibnamefont {Jiang}}, \bibinfo {author}
  {\bibfnamefont {D.}~\bibnamefont {Budker}}, \bibinfo {author} {\bibfnamefont
  {P.~R.}\ \bibnamefont {Hemmer}}, \bibinfo {author} {\bibfnamefont
  {A.}~\bibnamefont {Yacoby}}, \bibinfo {author} {\bibfnamefont
  {R.}~\bibnamefont {Walsworth}}, \ and\ \bibinfo {author} {\bibfnamefont
  {M.~D.}\ \bibnamefont {Lukin}},\ }\href {\doibase 10.1038/nphys1075}
  {\bibfield  {journal} {\bibinfo  {journal} {Nat. Phys.}\ }\textbf {\bibinfo
  {volume} {4}},\ \bibinfo {pages} {810} (\bibinfo {year} {2008})},\ \Eprint
  {http://arxiv.org/abs/0805.1367} {arXiv:0805.1367} \BibitemShut {NoStop}%
\bibitem [{\citenamefont {Mittermaier}\ and\ \citenamefont
  {Kay}(2009)}]{Mittermaier2009}%
  \BibitemOpen
  \bibfield  {author} {\bibinfo {author} {\bibfnamefont {A.~K.}\ \bibnamefont
  {Mittermaier}}\ and\ \bibinfo {author} {\bibfnamefont {L.~E.}\ \bibnamefont
  {Kay}},\ }\href {\doibase 10.1016/j.tibs.2009.07.004} {\bibfield  {journal}
  {\bibinfo  {journal} {Trends Biochem. Sci.}\ }\textbf {\bibinfo {volume}
  {34}},\ \bibinfo {pages} {601} (\bibinfo {year} {2009})}\BibitemShut
  {NoStop}%
\bibitem [{\citenamefont {Zhong}\ and\ \citenamefont {Xin}(2001)}]{Zhong2001}%
  \BibitemOpen
  \bibfield  {author} {\bibinfo {author} {\bibfnamefont {S.}~\bibnamefont
  {Zhong}}\ and\ \bibinfo {author} {\bibfnamefont {H.}~\bibnamefont {Xin}},\
  }\href {\doibase 10.1016/S0009-2614(00)01343-9} {\bibfield  {journal}
  {\bibinfo  {journal} {Chem. Phys. Lett.}\ }\textbf {\bibinfo {volume}
  {333}},\ \bibinfo {pages} {133} (\bibinfo {year} {2001})}\BibitemShut
  {NoStop}%
\bibitem [{\citenamefont {Bylander}\ \emph {et~al.}(2011)\citenamefont
  {Bylander}, \citenamefont {Gustavsson}, \citenamefont {Yan}, \citenamefont
  {Yoshihara}, \citenamefont {Harrabi}, \citenamefont {Fitch}, \citenamefont
  {Cory}, \citenamefont {Nakamura}, \citenamefont {Tsai},\ and\ \citenamefont
  {Oliver}}]{Bylander2011}%
  \BibitemOpen
  \bibfield  {author} {\bibinfo {author} {\bibfnamefont {J.}~\bibnamefont
  {Bylander}}, \bibinfo {author} {\bibfnamefont {S.}~\bibnamefont
  {Gustavsson}}, \bibinfo {author} {\bibfnamefont {F.}~\bibnamefont {Yan}},
  \bibinfo {author} {\bibfnamefont {F.}~\bibnamefont {Yoshihara}}, \bibinfo
  {author} {\bibfnamefont {K.}~\bibnamefont {Harrabi}}, \bibinfo {author}
  {\bibfnamefont {G.}~\bibnamefont {Fitch}}, \bibinfo {author} {\bibfnamefont
  {D.~G.}\ \bibnamefont {Cory}}, \bibinfo {author} {\bibfnamefont
  {Y.}~\bibnamefont {Nakamura}}, \bibinfo {author} {\bibfnamefont {J.-S.}\
  \bibnamefont {Tsai}}, \ and\ \bibinfo {author} {\bibfnamefont {W.~D.}\
  \bibnamefont {Oliver}},\ }\href {\doibase 10.1038/nphys1994} {\bibfield
  {journal} {\bibinfo  {journal} {Nat. Phys.}\ }\textbf {\bibinfo {volume}
  {7}},\ \bibinfo {pages} {565} (\bibinfo {year} {2011})}\BibitemShut {NoStop}%
\bibitem [{\citenamefont {Zhang}\ \emph {et~al.}(2007)\citenamefont {Zhang},
  \citenamefont {Peng}, \citenamefont {Rajendran},\ and\ \citenamefont
  {Suter}}]{Zhang2007}%
  \BibitemOpen
  \bibfield  {author} {\bibinfo {author} {\bibfnamefont {J.}~\bibnamefont
  {Zhang}}, \bibinfo {author} {\bibfnamefont {X.}~\bibnamefont {Peng}},
  \bibinfo {author} {\bibfnamefont {N.}~\bibnamefont {Rajendran}}, \ and\
  \bibinfo {author} {\bibfnamefont {D.}~\bibnamefont {Suter}},\ }\href
  {\doibase 10.1103/PhysRevA.75.042314} {\bibfield  {journal} {\bibinfo
  {journal} {Phys. Rev. A}\ }\textbf {\bibinfo {volume} {75}},\ \bibinfo
  {pages} {042314} (\bibinfo {year} {2007})}\BibitemShut {NoStop}%
\bibitem [{\citenamefont {{Ali Ahmed}}\ \emph {et~al.}(2013)\citenamefont {{Ali
  Ahmed}}, \citenamefont {{\'{A}}lvarez},\ and\ \citenamefont
  {Suter}}]{AliAhmed2013}%
  \BibitemOpen
  \bibfield  {author} {\bibinfo {author} {\bibfnamefont {M.}~\bibnamefont {{Ali
  Ahmed}}}, \bibinfo {author} {\bibfnamefont {G.}~\bibnamefont
  {{\'{A}}lvarez}}, \ and\ \bibinfo {author} {\bibfnamefont {D.}~\bibnamefont
  {Suter}},\ }\href {\doibase 10.1103/PhysRevA.87.042309} {\bibfield  {journal}
  {\bibinfo  {journal} {Phys. Rev. A}\ }\textbf {\bibinfo {volume} {87}},\
  \bibinfo {pages} {042309} (\bibinfo {year} {2013})}\BibitemShut {NoStop}%
\bibitem [{\citenamefont {Almog}\ \emph {et~al.}(2011)\citenamefont {Almog},
  \citenamefont {Sagi}, \citenamefont {Gordon}, \citenamefont {Bensky},
  \citenamefont {Kurizki},\ and\ \citenamefont {Davidson}}]{Almog2011}%
  \BibitemOpen
  \bibfield  {author} {\bibinfo {author} {\bibfnamefont {I.}~\bibnamefont
  {Almog}}, \bibinfo {author} {\bibfnamefont {Y.}~\bibnamefont {Sagi}},
  \bibinfo {author} {\bibfnamefont {G.}~\bibnamefont {Gordon}}, \bibinfo
  {author} {\bibfnamefont {G.}~\bibnamefont {Bensky}}, \bibinfo {author}
  {\bibfnamefont {G.}~\bibnamefont {Kurizki}}, \ and\ \bibinfo {author}
  {\bibfnamefont {N.}~\bibnamefont {Davidson}},\ }\href {\doibase
  10.1088/0953-4075/44/15/154006} {\bibfield  {journal} {\bibinfo  {journal}
  {J. Phys. B}\ }\textbf {\bibinfo {volume} {44}},\ \bibinfo {pages} {154006}
  (\bibinfo {year} {2011})}\BibitemShut {NoStop}%
\bibitem [{\citenamefont {Banchi}\ \emph {et~al.}(2014)\citenamefont {Banchi},
  \citenamefont {Giorda},\ and\ \citenamefont {Zanardi}}]{Banchi2014}%
  \BibitemOpen
  \bibfield  {author} {\bibinfo {author} {\bibfnamefont {L.}~\bibnamefont
  {Banchi}}, \bibinfo {author} {\bibfnamefont {P.}~\bibnamefont {Giorda}}, \
  and\ \bibinfo {author} {\bibfnamefont {P.}~\bibnamefont {Zanardi}},\ }\href
  {\doibase 10.1103/PhysRevE.89.022102} {\bibfield  {journal} {\bibinfo
  {journal} {Phys. Rev. E}\ }\textbf {\bibinfo {volume} {89}},\ \bibinfo
  {pages} {022102} (\bibinfo {year} {2014})}\BibitemShut {NoStop}%
\bibitem [{\citenamefont {Vasile}\ \emph {et~al.}(2014)\citenamefont {Vasile},
  \citenamefont {Galve},\ and\ \citenamefont {Zambrini}}]{Vasile2014}%
  \BibitemOpen
  \bibfield  {author} {\bibinfo {author} {\bibfnamefont {R.}~\bibnamefont
  {Vasile}}, \bibinfo {author} {\bibfnamefont {F.}~\bibnamefont {Galve}}, \
  and\ \bibinfo {author} {\bibfnamefont {R.}~\bibnamefont {Zambrini}},\ }\href
  {\doibase 10.1103/PhysRevA.89.022109} {\bibfield  {journal} {\bibinfo
  {journal} {Phys. Rev. A}\ }\textbf {\bibinfo {volume} {89}},\ \bibinfo
  {pages} {022109} (\bibinfo {year} {2014})}\BibitemShut {NoStop}%
\bibitem [{\citenamefont {Chin}\ \emph {et~al.}(2012)\citenamefont {Chin},
  \citenamefont {Huelga},\ and\ \citenamefont {Plenio}}]{Chin2012}%
  \BibitemOpen
  \bibfield  {author} {\bibinfo {author} {\bibfnamefont {A.~W.}\ \bibnamefont
  {Chin}}, \bibinfo {author} {\bibfnamefont {S.~F.}\ \bibnamefont {Huelga}}, \
  and\ \bibinfo {author} {\bibfnamefont {M.~B.}\ \bibnamefont {Plenio}},\
  }\href {\doibase 10.1103/PhysRevLett.109.233601} {\bibfield  {journal}
  {\bibinfo  {journal} {Phys. Rev. Lett.}\ }\textbf {\bibinfo {volume} {109}},\
  \bibinfo {pages} {233601} (\bibinfo {year} {2012})}\BibitemShut {NoStop}%
\bibitem [{\citenamefont {Haikka}\ and\ \citenamefont
  {Maniscalco}(2014)}]{Haikka2014}%
  \BibitemOpen
  \bibfield  {author} {\bibinfo {author} {\bibfnamefont {P.}~\bibnamefont
  {Haikka}}\ and\ \bibinfo {author} {\bibfnamefont {S.}~\bibnamefont
  {Maniscalco}},\ }\href {\doibase 10.1142/S1230161214400058} {\bibfield
  {journal} {\bibinfo  {journal} {Open Syst. Inf. Dyn.}\ }\textbf {\bibinfo
  {volume} {21}},\ \bibinfo {pages} {1440005} (\bibinfo {year}
  {2014})}\BibitemShut {NoStop}%
\bibitem [{\citenamefont {Paris}(2009)}]{Paris2009}%
  \BibitemOpen
  \bibfield  {author} {\bibinfo {author} {\bibfnamefont {M.~G.~A.}\
  \bibnamefont {Paris}},\ }\href {\doibase 10.1142/S0219749909004839}
  {\bibfield  {journal} {\bibinfo  {journal} {Int. J. Quantum Inf.}\ }\textbf
  {\bibinfo {volume} {7}},\ \bibinfo {pages} {125} (\bibinfo {year}
  {2009})}\BibitemShut {NoStop}%
\bibitem [{\citenamefont {{\'{A}}lvarez}\ and\ \citenamefont
  {Suter}(2011)}]{Alvarez2011}%
  \BibitemOpen
  \bibfield  {author} {\bibinfo {author} {\bibfnamefont {G.~A.}\ \bibnamefont
  {{\'{A}}lvarez}}\ and\ \bibinfo {author} {\bibfnamefont {D.}~\bibnamefont
  {Suter}},\ }\href {\doibase 10.1103/PhysRevLett.107.230501} {\bibfield
  {journal} {\bibinfo  {journal} {Phys. Rev. Lett.}\ }\textbf {\bibinfo
  {volume} {107}},\ \bibinfo {pages} {230501} (\bibinfo {year}
  {2011})}\BibitemShut {NoStop}%
\bibitem [{\citenamefont {Paris}(2014)}]{Paris2014}%
  \BibitemOpen
  \bibfield  {author} {\bibinfo {author} {\bibfnamefont {M.~G.~A.}\
  \bibnamefont {Paris}},\ }\href {\doibase 10.1016/j.physa.2014.06.052}
  {\bibfield  {journal} {\bibinfo  {journal} {Phys. A Stat. Mech. its Appl.}\
  }\textbf {\bibinfo {volume} {413}},\ \bibinfo {pages} {256} (\bibinfo {year}
  {2014})}\BibitemShut {NoStop}%
\bibitem [{\citenamefont {Benedetti}\ \emph
  {et~al.}(2014{\natexlab{a}})\citenamefont {Benedetti}, \citenamefont
  {Buscemi}, \citenamefont {Bordone},\ and\ \citenamefont
  {Paris}}]{Benedetti2014a}%
  \BibitemOpen
  \bibfield  {author} {\bibinfo {author} {\bibfnamefont {C.}~\bibnamefont
  {Benedetti}}, \bibinfo {author} {\bibfnamefont {F.}~\bibnamefont {Buscemi}},
  \bibinfo {author} {\bibfnamefont {P.}~\bibnamefont {Bordone}}, \ and\
  \bibinfo {author} {\bibfnamefont {M.~G.~A.}\ \bibnamefont {Paris}},\ }\href
  {\doibase 10.1103/PhysRevA.89.032114} {\bibfield  {journal} {\bibinfo
  {journal} {Phys. Rev. A}\ }\textbf {\bibinfo {volume} {89}},\ \bibinfo
  {pages} {032114} (\bibinfo {year} {2014}{\natexlab{a}})}\BibitemShut
  {NoStop}%
\bibitem [{\citenamefont {Cywi{\'{n}}ski}(2014)}]{Cywinski2014}%
  \BibitemOpen
  \bibfield  {author} {\bibinfo {author} {\bibfnamefont {L.}~\bibnamefont
  {Cywi{\'{n}}ski}},\ }\href {\doibase 10.1103/PhysRevA.90.042307} {\bibfield
  {journal} {\bibinfo  {journal} {Phys. Rev. A}\ }\textbf {\bibinfo {volume}
  {90}},\ \bibinfo {pages} {042307} (\bibinfo {year} {2014})}\BibitemShut
  {NoStop}%
\bibitem [{\citenamefont {Benedetti}\ and\ \citenamefont
  {Paris}(2014)}]{Benedetti2014b}%
  \BibitemOpen
  \bibfield  {author} {\bibinfo {author} {\bibfnamefont {C.}~\bibnamefont
  {Benedetti}}\ and\ \bibinfo {author} {\bibfnamefont {M.~G.~A.}\ \bibnamefont
  {Paris}},\ }\href {\doibase http://dx.doi.org/10.1016/j.physleta.2014.06.043}
  {\bibfield  {journal} {\bibinfo  {journal} {Phys. Lett. A}\ }\textbf
  {\bibinfo {volume} {378}},\ \bibinfo {pages} {2495 } (\bibinfo {year}
  {2014})}\BibitemShut {NoStop}%
\bibitem [{\citenamefont {Crow}\ and\ \citenamefont {Joynt}(2014)}]{Crow2014}%
  \BibitemOpen
  \bibfield  {author} {\bibinfo {author} {\bibfnamefont {D.}~\bibnamefont
  {Crow}}\ and\ \bibinfo {author} {\bibfnamefont {R.}~\bibnamefont {Joynt}},\
  }\href {\doibase 10.1103/PhysRevA.89.042123} {\bibfield  {journal} {\bibinfo
  {journal} {Phys. Rev. A}\ }\textbf {\bibinfo {volume} {89}},\ \bibinfo
  {pages} {042123} (\bibinfo {year} {2014})}\BibitemShut {NoStop}%
\bibitem [{\citenamefont {Neder}\ \emph {et~al.}(2011)\citenamefont {Neder},
  \citenamefont {Rudner}, \citenamefont {Bluhm}, \citenamefont {Foletti},
  \citenamefont {Halperin},\ and\ \citenamefont {Yacoby}}]{Neder2011}%
  \BibitemOpen
  \bibfield  {author} {\bibinfo {author} {\bibfnamefont {I.}~\bibnamefont
  {Neder}}, \bibinfo {author} {\bibfnamefont {M.~S.}\ \bibnamefont {Rudner}},
  \bibinfo {author} {\bibfnamefont {H.}~\bibnamefont {Bluhm}}, \bibinfo
  {author} {\bibfnamefont {S.}~\bibnamefont {Foletti}}, \bibinfo {author}
  {\bibfnamefont {B.~I.}\ \bibnamefont {Halperin}}, \ and\ \bibinfo {author}
  {\bibfnamefont {A.}~\bibnamefont {Yacoby}},\ }\href {\doibase
  10.1103/PhysRevB.84.035441} {\bibfield  {journal} {\bibinfo  {journal} {Phys.
  Rev. B}\ }\textbf {\bibinfo {volume} {84}},\ \bibinfo {pages} {035441}
  (\bibinfo {year} {2011})},\ \Eprint {http://arxiv.org/abs/1103.4862}
  {arXiv:1103.4862} \BibitemShut {NoStop}%
\bibitem [{\citenamefont {Biercuk}\ and\ \citenamefont
  {Bluhm}(2011)}]{Biercuk2011}%
  \BibitemOpen
  \bibfield  {author} {\bibinfo {author} {\bibfnamefont {M.~J.}\ \bibnamefont
  {Biercuk}}\ and\ \bibinfo {author} {\bibfnamefont {H.}~\bibnamefont
  {Bluhm}},\ }\href {\doibase 10.1103/PhysRevB.83.235316} {\bibfield  {journal}
  {\bibinfo  {journal} {Phys. Rev. B}\ }\textbf {\bibinfo {volume} {83}},\
  \bibinfo {pages} {235316} (\bibinfo {year} {2011})},\ \Eprint
  {http://arxiv.org/abs/1101.5189} {arXiv:1101.5189} \BibitemShut {NoStop}%
\bibitem [{\citenamefont {Witzel}\ \emph {et~al.}(2014)\citenamefont {Witzel},
  \citenamefont {Young},\ and\ \citenamefont {{Das Sarma}}}]{Witzel2014}%
  \BibitemOpen
  \bibfield  {author} {\bibinfo {author} {\bibfnamefont {W.~M.}\ \bibnamefont
  {Witzel}}, \bibinfo {author} {\bibfnamefont {K.}~\bibnamefont {Young}}, \
  and\ \bibinfo {author} {\bibfnamefont {S.}~\bibnamefont {{Das Sarma}}},\
  }\href {\doibase 10.1103/PhysRevB.90.115431} {\bibfield  {journal} {\bibinfo
  {journal} {Phys. Rev. B}\ }\textbf {\bibinfo {volume} {90}},\ \bibinfo
  {pages} {115431} (\bibinfo {year} {2014})}\BibitemShut {NoStop}%
\bibitem [{\citenamefont {Fink}\ and\ \citenamefont {Bluhm}(2014)}]{Fink2014}%
  \BibitemOpen
  \bibfield  {author} {\bibinfo {author} {\bibfnamefont {T.}~\bibnamefont
  {Fink}}\ and\ \bibinfo {author} {\bibfnamefont {H.}~\bibnamefont {Bluhm}},\
  }\href {http://arxiv.org/abs/1402.0235} {\  (\bibinfo {year} {2014})},\
  \Eprint {http://arxiv.org/abs/1402.0235} {arXiv:1402.0235} \BibitemShut
  {NoStop}%
\bibitem [{\citenamefont {Yu}\ and\ \citenamefont {Eberly}(2010)}]{Yu2010}%
  \BibitemOpen
  \bibfield  {author} {\bibinfo {author} {\bibfnamefont {T.}~\bibnamefont
  {Yu}}\ and\ \bibinfo {author} {\bibfnamefont {J.}~\bibnamefont {Eberly}},\
  }\href {http://dx.doi.org/10.1016/j.optcom.2009.10.042
  http://www.sciencedirect.com/science/article/pii/S0030401809010347
  http://www.sciencedirect.com/science/article/pii/S0030401809010347/pdf?md5=e6bc12c8ee0c3de59ef61537a63fd597{\&}pid=1-s2.0-S0030401809010347-main}
  {\bibfield  {journal} {\bibinfo  {journal} {Opt. Comm.}\ }\textbf {\bibinfo
  {volume} {283}},\ \bibinfo {pages} {676} (\bibinfo {year}
  {2010})}\BibitemShut {NoStop}%
\bibitem [{\citenamefont {Li}\ and\ \citenamefont {Liang}(2011)}]{Li2011}%
  \BibitemOpen
  \bibfield  {author} {\bibinfo {author} {\bibfnamefont {J.-Q.}\ \bibnamefont
  {Li}}\ and\ \bibinfo {author} {\bibfnamefont {J.-Q.}\ \bibnamefont {Liang}},\
  }\href@noop {} {\bibfield  {journal} {\bibinfo  {journal} {Phys. Lett. A}\
  }\textbf {\bibinfo {volume} {375}},\ \bibinfo {pages} {1496} (\bibinfo {year}
  {2011})}\BibitemShut {NoStop}%
\bibitem [{\citenamefont {Benedetti}\ \emph {et~al.}(2013)\citenamefont
  {Benedetti}, \citenamefont {Buscemi}, \citenamefont {Bordone},\ and\
  \citenamefont {Paris}}]{Benedetti2013a}%
  \BibitemOpen
  \bibfield  {author} {\bibinfo {author} {\bibfnamefont {C.}~\bibnamefont
  {Benedetti}}, \bibinfo {author} {\bibfnamefont {F.}~\bibnamefont {Buscemi}},
  \bibinfo {author} {\bibfnamefont {P.}~\bibnamefont {Bordone}}, \ and\
  \bibinfo {author} {\bibfnamefont {M.~G.~A.}\ \bibnamefont {Paris}},\ }\href
  {\doibase 10.1103/PhysRevA.87.052328} {\bibfield  {journal} {\bibinfo
  {journal} {Phys. Rev. A}\ }\textbf {\bibinfo {volume} {87}},\ \bibinfo
  {pages} {052328} (\bibinfo {year} {2013})}\BibitemShut {NoStop}%
\bibitem [{\citenamefont {Benedetti}\ \emph {et~al.}(2012)\citenamefont
  {Benedetti}, \citenamefont {Buscemi}, \citenamefont {Bordone},\ and\
  \citenamefont {Paris}}]{Benedetti2012}%
  \BibitemOpen
  \bibfield  {author} {\bibinfo {author} {\bibfnamefont {C.}~\bibnamefont
  {Benedetti}}, \bibinfo {author} {\bibfnamefont {F.}~\bibnamefont {Buscemi}},
  \bibinfo {author} {\bibfnamefont {P.}~\bibnamefont {Bordone}}, \ and\
  \bibinfo {author} {\bibfnamefont {M.~G.~A.}\ \bibnamefont {Paris}},\ }\href
  {\doibase 10.1142/S0219749912410055} {\bibfield  {journal} {\bibinfo
  {journal} {Int. J. Quantum Inf.}\ }\textbf {\bibinfo {volume} {10}},\
  \bibinfo {pages} {1241005} (\bibinfo {year} {2012})}\BibitemShut {NoStop}%
\bibitem [{\citenamefont {Li}\ \emph {et~al.}(2013)\citenamefont {Li},
  \citenamefont {Silveri}, \citenamefont {Kumar}, \citenamefont {Pirkkalainen},
  \citenamefont {Veps{\"{a}}l{\"{a}}inen}, \citenamefont {Chien}, \citenamefont
  {Tuorila}, \citenamefont {Sillanp{\"{a}}{\"{a}}}, \citenamefont {Hakonen},
  \citenamefont {Thuneberg},\ and\ \citenamefont {Paraoanu}}]{Li2013}%
  \BibitemOpen
  \bibfield  {author} {\bibinfo {author} {\bibfnamefont {J.}~\bibnamefont
  {Li}}, \bibinfo {author} {\bibfnamefont {M.~P.}\ \bibnamefont {Silveri}},
  \bibinfo {author} {\bibfnamefont {K.~S.}\ \bibnamefont {Kumar}}, \bibinfo
  {author} {\bibfnamefont {J.-M.}\ \bibnamefont {Pirkkalainen}}, \bibinfo
  {author} {\bibfnamefont {A.}~\bibnamefont {Veps{\"{a}}l{\"{a}}inen}},
  \bibinfo {author} {\bibfnamefont {W.~C.}\ \bibnamefont {Chien}}, \bibinfo
  {author} {\bibfnamefont {J.}~\bibnamefont {Tuorila}}, \bibinfo {author}
  {\bibfnamefont {M.~a.}\ \bibnamefont {Sillanp{\"{a}}{\"{a}}}}, \bibinfo
  {author} {\bibfnamefont {P.~J.}\ \bibnamefont {Hakonen}}, \bibinfo {author}
  {\bibfnamefont {E.~V.}\ \bibnamefont {Thuneberg}}, \ and\ \bibinfo {author}
  {\bibfnamefont {G.~S.}\ \bibnamefont {Paraoanu}},\ }\href {\doibase
  10.1038/ncomms2383} {\bibfield  {journal} {\bibinfo  {journal} {Nat.
  Commun.}\ }\textbf {\bibinfo {volume} {4}},\ \bibinfo {pages} {1420}
  (\bibinfo {year} {2013})},\ \Eprint {http://arxiv.org/abs/1205.0675}
  {arXiv:1205.0675} \BibitemShut {NoStop}%
\bibitem [{\citenamefont {Burkard}(2009)}]{Burkard2009}%
  \BibitemOpen
  \bibfield  {author} {\bibinfo {author} {\bibfnamefont {G.}~\bibnamefont
  {Burkard}},\ }\href {\doibase 10.1103/PhysRevB.79.125317} {\bibfield
  {journal} {\bibinfo  {journal} {Phys. Rev. B}\ }\textbf {\bibinfo {volume}
  {79}},\ \bibinfo {pages} {1} (\bibinfo {year} {2009})},\ \Eprint
  {http://arxiv.org/abs/0803.0564} {arXiv:0803.0564} \BibitemShut {NoStop}%
\bibitem [{\citenamefont {Wold}\ \emph {et~al.}(2012)\citenamefont {Wold},
  \citenamefont {Brox}, \citenamefont {Galperin},\ and\ \citenamefont
  {Bergli}}]{Wold2012}%
  \BibitemOpen
  \bibfield  {author} {\bibinfo {author} {\bibfnamefont {H.~J.}\ \bibnamefont
  {Wold}}, \bibinfo {author} {\bibfnamefont {H.}~\bibnamefont {Brox}}, \bibinfo
  {author} {\bibfnamefont {Y.~M.}\ \bibnamefont {Galperin}}, \ and\ \bibinfo
  {author} {\bibfnamefont {J.}~\bibnamefont {Bergli}},\ }\href {\doibase
  10.1103/PhysRevB.86.205404} {\bibfield  {journal} {\bibinfo  {journal} {Phys.
  Rev. B}\ }\textbf {\bibinfo {volume} {86}},\ \bibinfo {pages} {205404}
  (\bibinfo {year} {2012})},\ \Eprint {http://arxiv.org/abs/1206.2174}
  {arXiv:1206.2174} \BibitemShut {NoStop}%
\bibitem [{\citenamefont {Benedetti}\ \emph
  {et~al.}(2014{\natexlab{b}})\citenamefont {Benedetti}, \citenamefont
  {Paris},\ and\ \citenamefont {Maniscalco}}]{Benedetti2014c}%
  \BibitemOpen
  \bibfield  {author} {\bibinfo {author} {\bibfnamefont {C.}~\bibnamefont
  {Benedetti}}, \bibinfo {author} {\bibfnamefont {M.~G.~A.}\ \bibnamefont
  {Paris}}, \ and\ \bibinfo {author} {\bibfnamefont {S.}~\bibnamefont
  {Maniscalco}},\ }\href {\doibase 10.1103/PhysRevA.89.012114} {\bibfield
  {journal} {\bibinfo  {journal} {Phys. Rev. A}\ }\textbf {\bibinfo {volume}
  {89}},\ \bibinfo {pages} {012114} (\bibinfo {year} {2014}{\natexlab{b}})},\
  \Eprint {http://arxiv.org/abs/1309.5270v1} {arXiv:1309.5270v1} \BibitemShut
  {NoStop}%
\bibitem [{\citenamefont {D'Ariano}\ \emph {et~al.}(2001)\citenamefont
  {D'Ariano}, \citenamefont {{Lo Presti}},\ and\ \citenamefont
  {Paris}}]{DAriano2001}%
  \BibitemOpen
  \bibfield  {author} {\bibinfo {author} {\bibfnamefont {G.~M.}\ \bibnamefont
  {D'Ariano}}, \bibinfo {author} {\bibfnamefont {P.}~\bibnamefont {{Lo
  Presti}}}, \ and\ \bibinfo {author} {\bibfnamefont {M.~G.~A.}\ \bibnamefont
  {Paris}},\ }\href {\doibase 10.1103/PhysRevLett.87.270404} {\bibfield
  {journal} {\bibinfo  {journal} {Phys. Rev. Lett.}\ }\textbf {\bibinfo
  {volume} {87}},\ \bibinfo {pages} {270404} (\bibinfo {year}
  {2001})}\BibitemShut {NoStop}%
\bibitem [{\citenamefont {Fujiwara}(2001)}]{Fujiwara2001}%
  \BibitemOpen
  \bibfield  {author} {\bibinfo {author} {\bibfnamefont {A.}~\bibnamefont
  {Fujiwara}},\ }\href {\doibase 10.1103/PhysRevA.63.042304} {\bibfield
  {journal} {\bibinfo  {journal} {Phys. Rev. A}\ }\textbf {\bibinfo {volume}
  {63}},\ \bibinfo {pages} {042304} (\bibinfo {year} {2001})}\BibitemShut
  {NoStop}%
\bibitem [{\citenamefont {Ac{\'{i}}n}\ \emph {et~al.}(2001)\citenamefont
  {Ac{\'{i}}n}, \citenamefont {Jan{\'{e}}},\ and\ \citenamefont
  {Vidal}}]{Acin2001}%
  \BibitemOpen
  \bibfield  {author} {\bibinfo {author} {\bibfnamefont {A.}~\bibnamefont
  {Ac{\'{i}}n}}, \bibinfo {author} {\bibfnamefont {E.}~\bibnamefont
  {Jan{\'{e}}}}, \ and\ \bibinfo {author} {\bibfnamefont {G.}~\bibnamefont
  {Vidal}},\ }\href {\doibase 10.1103/PhysRevA.64.050302} {\bibfield  {journal}
  {\bibinfo  {journal} {Phys. Rev. A}\ }\textbf {\bibinfo {volume} {64}},\
  \bibinfo {pages} {050302} (\bibinfo {year} {2001})}\BibitemShut {NoStop}%
\bibitem [{\citenamefont {Huelga}\ \emph {et~al.}(1997)\citenamefont {Huelga},
  \citenamefont {Macchiavello}, \citenamefont {Pellizzari}, \citenamefont
  {Ekert}, \citenamefont {Plenio},\ and\ \citenamefont {Cirac}}]{Huelga1997}%
  \BibitemOpen
  \bibfield  {author} {\bibinfo {author} {\bibfnamefont {S.~F.}\ \bibnamefont
  {Huelga}}, \bibinfo {author} {\bibfnamefont {C.}~\bibnamefont
  {Macchiavello}}, \bibinfo {author} {\bibfnamefont {T.}~\bibnamefont
  {Pellizzari}}, \bibinfo {author} {\bibfnamefont {A.~K.}\ \bibnamefont
  {Ekert}}, \bibinfo {author} {\bibfnamefont {M.~B.}\ \bibnamefont {Plenio}}, \
  and\ \bibinfo {author} {\bibfnamefont {J.~I.}\ \bibnamefont {Cirac}},\ }\href
  {\doibase 10.1103/PhysRevLett.79.3865} {\bibfield  {journal} {\bibinfo
  {journal} {Phys. Rev. Lett.}\ }\textbf {\bibinfo {volume} {79}},\ \bibinfo
  {pages} {3865} (\bibinfo {year} {1997})}\BibitemShut {NoStop}%
\end{thebibliography}%
\end{document}